# Direct and mediating influences of user-developer perception gaps in requirements understanding on user participation

Jingdong Jia (✉)

School of Software, Beihang University, No. 37 Xueyuan Road, Haidian District, Beijing, 100191, China

Tel.: +86 10 8231 6499

E-mail address: jiajingdong@buaa.edu.cn

Luiz Fernando Capretz

Department of Electrical & Computer Engineering, Western University, London, Ontario, N6A5B9, Canada.

Email address: lcapretz@uwo.ca

**Abstract** User participation is considered an effective way to conduct requirements engineering, but user-developer perception gaps in requirements understanding occur frequently. Since user participation in practice is not as active as we expect and the requirements perception gap has been recognized as a risk that negatively affects projects, exploring whether user-developer perception gaps in requirements understanding will hinder user participation is worthwhile. This will help develop a greater comprehension of the intertwined relationship between user participation and perception gap, a topic that has not yet been extensively examined. This study investigates the direct and mediating influences of user-developer requirements perception gaps on user participation by integrating requirements uncertainty and top management support. Survey data collected from 140 subjects were examined and analyzed using structural equation modeling. The results indicate that perception gaps have a direct negative effect on user participation and negate completely the positive effect of top management support on user participation. Additionally, perception gaps do not have a mediating effect between requirements uncertainty and user participation because requirements uncertainty does not significantly and directly affect user participation, but requirements uncertainty indirectly influences user participation due to its significant direct effect on perception gaps. The theoretical and practical implications are discussed, and limitations and possible future research areas are identified.

# 1 Introduction

Requirements engineering (RE) is a crucial step for any software project because incomplete and incorrect requirements inevitably propagate into the later stages of software development, leading to implementations that do not meet user needs and failure of the final project [1]. User participation is a well-established principle to facilitate requirements elicitation. It is widely accepted that user participation plays a positive key role in RE. However, some problems also occur along with user participation. One problem is perception gaps in requirements understanding between users and developers.

The user-developer perception gap in requirements understanding is a critical problem in RE [2, 3]. It has been regarded as one kind of risk of requirements determination [4, 5], and a leading reason that software projects are problematic or fail entirely [1]. Some authors have empirically validated that the user-developer perception gap in requirements understanding is negatively associated with project performance [4, 5]. Therefore, we wonder whether it has any direct or indirect negative effects on other aspects related to RE. In this paper, we focus on investigating its effects on user participation. For one thing, the user-developer perception gap in requirements understanding and user participation intertwine: the perception gap emerges in the process of user participation during requirements determination. The methods proposed to reduce this gap are inseparable from user participation [3, 4, 6], demonstrating that user participation has a positive effect on closing the gap. However, whether the reverse relationship from the user-developer perception gap in requirements understanding to user participation exists is still a question. In addition, although most literature finds that user participation has a positive impact on requirements elicitation and further project success, there are still some arguments about its negative impacts on system success [7-10]. And we often observe that user participation in practice is not as active as we expect in RE. Thus, examining which factors impede user participation to achieve the desired effect is a worthwhile project. Some challenges and barriers that user participation faces have been identified [11], but the perception gap was not included in that research.

In addition to exploring whether user-developer perception gaps in requirements understanding directly and negatively influence user participation, we also want to examine whether perception gaps weaken or aggravate the positive or negative effects of other external factors on user participation. With regard to the latter, this study incorporates top management support and requirements uncertainty, because the former is related to human behavior and the latter is a non-human matter.

Top management is another critical stakeholder besides users, and the fact that their support has a positive effect on user participation has been validated [12, 13]. Our research question is whether the user-developer requirements perception gaps will weaken or even offset the positive effect of top management support on user participation. With regard to requirements uncertainty, it is an important environmental factor in RE. A considerable amount of empirical research supports that requirements uncertainty has a negative effect on project performance [14, 15], and user participation is necessary in order to reduce the extent of requirements uncertainty [16]. However, whether requirements uncertainty directly impedes the practice of user participation has not been empirically discussed. Given the negative relationship that exists, the question is whether user-developer perception gaps will deteriorate the negative influence of requirements uncertainty on user participation.

In summary, the primary purpose of this paper is to empirically assess the direct and mediating influences of user-developer perception gaps in requirements understanding on user participation. The analysis in this paper provides an opportunity not only to deeply recognize the danger of perception gaps as a risk in a project, but also to fully understand the interaction between gap and participation. Thus, this paper furthers the understanding of user participation problems and presents challenges to improve user participation practice in RE. Five hypotheses were proposed and assessed by the method of structural equation modeling.

This rest of the paper is organized as follows. Section 2 provides the theoretical background and hypotheses development. In Section 3, we describe our research method. The results of the analysis are presented in Section 4. We subsequently discuss the results in terms of their theoretical and practical contributions in Section 5. Finally, we conclude this paper by specifying the limitations of this work and making suggestions for future research.

**2 Theoretical background and hypotheses development**

2.1 User participation and perception gap

The terms "user participation" and "user involvement" are very similar; the former refers to behaviors and activities and the latter refers to the psychological state of the user [17]. However, they are often used interchangeably, and the fact that they have a positive influence on each other has been validated [18], so in the paper we employ the term "user participation" to represent a comprehensive user practice, including psychological state and actual behavior. User participation is a widely accepted practice, especially in agile RE; users are treated as team members, collocated with the team, and available to provide information and discuss the project issues with developers [19]. However, user participation also faces some challenges and barriers [11].

Perception is viewed as a cognitive process in which individuals notice, encode, store, and retrieve information about the world around them. Based on experience, background, training, and environment, people gradually and continually develop their own schema for understanding the world. Once developed, these schemas shape our perceptions of virtually everything. Formally, perception gap is defined as the existence of multiple and conflicting interpretations about an organizational situation by different stakeholders [4, 5]. Perception gaps are large when frames of reference differ [20]. Because users and developers have different knowledge, interests, expectations, and problem-solving approaches, they often exhibit completely different frames of reference when interpreting information [4]. Additionally, users might have more interest in how the system will improve their jobs, but the developers' main interests might be in how well they will do their jobs during RE [21]. Thus, unsurprisingly, user-developer perception gaps in requirements understanding occur.

Perception gaps in requirements understanding raise barriers to software development through communication difficulties and misunderstandings [7, 22], thus leading to all kinds of conflicts. Yeh and Tsai [23] argued that there were two conflict potentials between users and developers: physical and mental levels. From the mental viewpoint, the larger the perception gap is, the stronger will be the frustrations of the users. Users think their views are missing and that developers do not take user opinions and experiences into consideration or are not interested in user feedback when developing software. Users even think that developers view themselves much more positively than they view users. According to a recent investigation in [8], most users disagreed with the statement that "software providers are interested in end-users' feedback." The investigation result is in line with an early study by Hunton [24] who discovered that there was a high degree of user desire to become involved in the software development process and a low level of actual participation. The incongruence between these two participation constructs can be attributed to the frustrated feeling users have of the lack of concern by software developers. Additionally, when facing conflicts, some people may take an avoiding strategy [25, 26]. This also implies that users possibly decrease participation behaviors through avoiding future cooperation or interaction with developers due to conflicts brought on by perception gaps in requirements understanding. Thus, a good portion of the decrease of user participation in a project is explained by perception gaps. So, we hypothesize the following:

**H1:** User-developer perception gaps have a negative effect on user participation.

2.2 Requirements uncertainty and perception gap

Requirements uncertainty is a major risk in RE. Three distinct dimensions of requirements uncertainty have been observed: (1) requirements instability – the extent of changes in user requirements over the course of the project, (2) user uncertainty – the ability of users to specify requirements, and (3) analyst uncertainty – the ability of analysts to elicit and evaluate requirements [4]. The latter two dimensions are also referred as requirements diversity representing the extent to which stakeholders differ in their views of system requirements [27]. Therefore, the more requirements diversity, the more perception gaps among stakeholders.

Sometimes, requirements instability is necessary from a managerial perspective because changing system requirements may be available options for capitalizing on business opportunities. Recognizing the need to make changes is one thing, however, incorporating those changes in the requirements of an ongoing software project is another [27]. Software developers are usually worried about requirements uncertainty because it makes understanding the project requirements and scope more difficult. Requirements instability can potentially exacerbate problems associated with interpersonal conflict, such as disagreements on project goals and the hidden opinions of divergent stakeholders [27, 28].

With greater uncertainty, achieving "consonance" on requirements between users and developers becomes more difficult. When there is a greater volatility in requirements and a greater extent of difference among stakeholders about the requirements, the complexity of determining system requirements increases and this increases the perception gaps among stakeholders [5]. Thus, requirements uncertainty is one source of stakeholder perception gaps in requirements understanding. Perception gaps might incur further requirements changes [29]. Although they are intertwined, we are concerned about the link from requirements uncertainty to perception gaps, which was empirically supported in [5]. Therefore,

**H2:** Requirements uncertainty has a positive effect on user-developer perception gaps.

2.3 Requirements uncertainty and user participation

Requirements change is unavoidable, and can be seen as an inherent attribute of a software project. Thus, decades of research have been devoted to investigating which methods can reduce requirements uncertainty. User participation is one highly recommended way to reduce requirements uncertainty [10, 30, 31]. Yet, things generally interact with each other, so there is a reason to doubt whether requirements uncertainty actually affects user participation.

Recently, Bano and Zowghi [9] thought that with higher requirements uncertainty, higher user participation was required. This opinion is similar to that of Emam et al. [32]: the importance of user participation should increase as uncertainty increases. However, the importance of user participation or the required user participation is different from the actual user participation. In fact, these opinions tended to emphasize the fact that user participation is necessary to cope with requirements uncertainty, rather than the impact of requirements uncertainty on the actual user participation.

Sometimes requirements changes come from the organizational environment, in which case resistant users do not welcome these changes for different reasons. First, user horizons are limited by their roles and responsibility, so users may not have a global view to understand the necessity of change. Generally, users think requirements analysis is extra work because it is beyond normal business. Requirements uncertainty will make users feel that they have to make greater effort. Consequently, excessive energy consumption will decrease user willingness to participate. Second, some changes would not coincide with user interests. For example, business process changes will increase user workload or reduce user decision-making powers. Under this circumstance, expecting users to participate and help developers analyze requirements is impossible. Third, frequent requirements changes will make users feel that their previous work is for nothing and they will begin to doubt the value of their participation. Therefore, it is possible for users not to embrace some requirements uncertainty, further decreasing the participation willingness or behavior. Additionally, because uncertainty is one of the challenges of user participation [9], uncertainty may impede actual user participation. From the above discussion we expect that:

**H3:** Requirements uncertainty has a negative effect on user participation.

2.4 Top management support, perception gap, and user participation

Top management support has a significant effect on software project success [33, 34]. In contrast, lack of such support is a major impediment to the favorable outcome of software projects [12, 35-37]. Why top management support is so important for software projects is easy to understand. First, top managers, with their broader perspectives, are in a better position than developers to identify business opportunities for the exploitation of software projects [38], so they can help developers elicit requirements more integrally. Second, a software development project usually involves huge investments and often requires organization-wide business change; top management support can promote resource allocation and organizational change implementation. Third, top management support sends a clear message to all the stakeholders regarding the importance of the software project. Top management also plays an important role in creating and fostering an organizational climate to encourage and motivate their employees to invest their best efforts in their work [13].

For users, visible top management support, for example, attending or hosting a project meeting, encourages positive user attitudes toward software development and leads to a smoother conversion from the existing work procedures. It may also lead to more positive participatory behavior [38], for example, putting more time and energy into the project and helping define project requirements. Lack of top management support was seen as one of the originating reasons for user participation problems [9]. Therefore, top management support plays a positive and direct role on user participation, which is supported in [13].

In addition, top management has the authority to influence other related stakeholders through negotiation, persuasion, and resource provision and also by motivating parties to cooperate with each other [39], which provide impetus for developers to be involved in collaboration with users. When a big conflict happens between users and developers, it is not hard to see that the question is often solved by top management in practice. Essentially, one of the major functions of top management support is to build an idea and culture to ensure project success. Therefore, users and developers will try to seek concurrence and reduce differences during requirements analysis. They both will develop and enhance a more thorough and coincident requirements understanding, which in turn stimulates the decrease of requirements perception gaps between them. Chen et al. [4] have stated that top management support, a major pre-project partnering activity, has a significant positive effect on closing the perception gaps between users and developers.

Based on the above discussion and empirical evidence in the literature, the following hypotheses are proposed:

**H4:** Top management support has a positive effect on user participation.

**H5:** Top management support has a positive effect on reducing the perception gap.

The five stated hypotheses will be tested to examine whether the user-developer perception gap has a direct negative influence on user participation and whether the perception gap weakens the positive impact of top management support or aggravates the negative impact of requirements uncertainty on user participation.

## 3 Research Method

### 3.1 Development of instrument

A survey was employed to test the research hypotheses. Compared with case studies and experiments, this kind of empirical research method tend to cover more subjects. Each construct was measured by multiple survey items to ensure reliability. All measurement items were derived from relevant prior research, then developed and refined based on the context of this study.

Five question items of the requirements uncertainty construct were drawn from the references of [15, 40]. Perception gap was measured using five items developed by Chen et al. [4]. All of the above items were anchored on a Likert scale from 1 (strongly disagree) to 5 (strongly agree). User participation was measured using a five-point scale ranging from very little (1) to very much (5), and six relevant items were extracted from the studies of [41-43]. For top management support, using the questions established in [38, 44-46], we developed a Likert five-point scale from 1 (totally disagree) to 5 (totally agree) of six items. A full list of the measurement items is provided in Appendix A.

An English version survey was first developed and subsequently translated into Chinese. Then, in order to ensure that subjects could understand the questionnaire, it was piloted on fifteen students with experience in software projects.

Their feedback resulted in minor changes to the survey instructions, refinement in the wording of several items, and an additional explanation of some terms.

3.2 Sample

The ideal candidate for our study was an experienced software developer or user. For practical reasons, subjects were part-time graduate students in a software school of a Chinese university. They were professionals familiar with the software project activities and had some work experience in a software company or had work experience closely related to software application in a non-software industry company. Therefore, the sample represents software developers and users. Paper questionnaires were distributed to students before a class and collected after the class, and electronic questionnaires were sent to students who were absent from the class and returned by an instant messenger software tool. In both cases, students were informed that the survey was intended for academic purposes and they voluntarily choose whether to participate in the survey or not. Additionally, they were assured that their responses would be kept confidential. Of the 170 questionnaires distributed, 152 were returned, yielding a response rate of 89.41%. Among the responses, 12 had identical answers to all the survey items and were therefore not used. As a result, 140 surveys were analyzed, giving a usable return rate of 92.1%.

Non-response bias occurs when the opinions and perceptions of the survey respondents do not accurately represent the entire sample to whom the survey was sent [47]. One test for non-response bias involves comparing the demographics of early versus late respondents. The characteristics of the 116 individuals who responded to the paper survey were compared to those of the 24 who responded to the electronic survey. No significant differences were found; therefore, all respondents were combined for further analysis.

Table 1 presents the demographic features of the sample. The characteristics of the organizations and projects that respondents serve are summarized in Appendix B. In Table 1, individual identity is related to the type of company in which individuals work: "software client" means that respondents work in non-software companies that purchase software services, also known as the first party, and "software provider" means that respondents work in software companies that supply software services, also known as the second party. Respondents were both software providers (75%) and software clients (24.3%). It seems that the percentage of users is low; however, in fact, some developers also play the role of users because they should put forward requirements for the software used within their company. It should be noted that the first four job positions exist in both the first party and the second party, thus two corresponding numbers were given for each job position; the job position of user comes from the first party, and the other job positions come from the second party.

Table 1 Demographic information

| Variables | Categories | | Number | Percent |
|---|---|---|---|---|
| Gender | Male | | 103 | 73.6 |
| | Female | | 37 | 26.4 |
| Individual identity | Software client (the first party) | | 34 | 24.3 |
| | Software provider (the second party) | | 105 | 75 |
| | Missing | | 1 | 0.7 |
| Job position | Top manager | the first party | 2 | 1.43 |
| | | the second party | 1 | 0.71 |
| | Business supervisor | the first party | 10 | 7.14 |
| | | the second party | 2 | 1.43 |
| | IT supervisor | the first party | 3 | 2.14 |
| | | the second party | 1 | 0.71 |
| | Software project manager | the first party | 10 | 7.14 |
| | | the second party | 20 | 14.29 |
| | User | | 9 | 6.42 |
| | Developer | | 54 | 38.6 |
| | Tester | | 21 | 15 |
| | Requirements analyst | | 6 | 4.28 |
| | Other (maintenance) | | 1 | 0.71 |
| Age | 20-25 | | 74 | 52.9 |
| | 26-30 | | 39 | 27.9 |
| | 31-35 | | 23 | 16.4 |
| | >=36 | | 3 | 2.1 |
| | missing | | 1 | 0.7 |

In Table 1, we can see that the major role of respondents is developer (38.6%), and the requirements analysts who actually do the activities of RE only account for 4.28%. However, according to the verbal communication between the first author and students in class, the work of some developers is actually related to RE for several reasons. First, the roles of software developers and requirements analysts are integrated in a small software company due to the lack of manpower. In this case, they prefer to view themselves as developers rather than analysts. Second, some developers serve in a company providing general software products, for example Weibo (a famous and influential Chinese social network product). Under this circumstance, developers need to discuss requirements with product managers who are also seen as users. Third, some development teams use the agile method, so developers necessarily participate in RE. Additionally, as mentioned before, some developers also act as users due to providing requirements for software used internally. In addition, we also can see that software project manager and tester account for the second and third roles of the sample. Although they may not directly do activities of RE, they need to learn the requirements during RE to manage a project or develop test cases. Thus, their responses can also provide useful data for the study. Additionally, we can see that only three subjects were top managers. Although the percentage is relatively low, it is in line with the fact that top managers do not have much time to pursue part-time studies because of their workload. However, software projects also need the support from department managers. Thus, business and IT supervisors can also be seen as top management. In our case, top management accounts for 13.6% of the sample. Therefore, the results were beneficial in investigating the user-developer perception gaps in requirements understanding and top management support.

**4 Data analysis and results**

To empirically test the hypotheses, we used structural equation modeling (SEM) and Amos 17.0, which is a good SEM software. SEM is a confirmatory method as it is intended to validate that the hypothetical relations among the latent (unobservable) variables and relationships between the latent and manifest (observed) variables are in accordance with obtained empirical data. It is, therefore, appropriate for analyzing theoretical models or research designs [48]. It has been widely used in empirical scientific research, especially in the social sciences [45]. Following the details of the SEM process described in [49], the measurement and structural models were assessed to ensure that the results were acceptable and were consistent with the underlying research hypotheses [46]. The measurement model was first estimated using confirmatory factor analysis (CFA) to test the overall fit, reliability, and validity of the constructs. Then, the hypotheses were examined through a structural model analysis.

4.1 Measurement model

Firstly, the overall model fit was assessed in order to determine the consistency level of a model as a whole with the available empirical data [45]. Several indices have been developed to measure the overall fit; each has a different meaning and performs differently. However, there is no agreement on which indices should be more applicable. We included the following fit indices that are most commonly used in the literature: the goodness-of-fit index (GFI), the adjusted goodness-of-fit index (AGFI), the Tucker–Lewis coefficient (TLI), the comparative fit index (CFI), the root mean squared residual (RMR), the root mean square error of approximation (RMSEA), and the ratio of $\chi^2$ to the degrees of freedom ($\chi^2$/df). Table 2 gives the recommended acceptance threshold value of each index; these were taken from the studies of [50, 51].

**Table 2** Goodness-of-fit indices for the measurement model

| Index | Rec. criteria | Initial model | Revised measurement model |
|---|---|---|---|
| GFI | ≥0.90 | *0.85* | 0.93 |
| AGFI | ≥0.80 | 0.81 | 0.89 |
| TLI | ≥0.90 | *0.87* | 0.94 |
| CFI | ≥0.90 | *0.88* | 0.98 |
| RMR | ≤0.10 | 0.07 | 0.06 |
| RMSEA | ≤0.08 | 0.06 | 0.06 |
| $\chi^2$/df | ≤3 | 1.42 | 1.44 |

As shown in Table 2, three indices of the initial measurement model, whose values are italicized, do not meet the recommended criteria, so the overall fit of the initial model is unsatisfactory. Thus, the initial model was revised to improve its goodness-of-fit with the data by dropping items based on factor loadings, the squared multiple correlations, and the modification indices. The modification is an iterative process; only one item was changed at a time. The revised measurement model exhibits a good fit because all its fit indices fell within the recommended criteria (Table 2). Items excluded from the revised measurement model are in italics in Appendix A. The remaining 11 items in the revised model were used for the subsequent analysis.

We next examined the reliability of the individual indicator by observing the factor loading of each indicator on its corresponding construct. To be viewed as having high reliability, factor loadings should be significant and not less than 0.50 [47, 52]. Table 3 indicates that all the factor loadings (ranging from 0.54 to 0.94) are statistically significant at the 0.05 level and meet the criteria. In addition, item-total correlation (ITC), which refers to the correlation between an individual indicator and the total score of all other indicators in the same construct, is also seen as a measure of the reliability of indicators (>.30 recommended) [27]. Table 3 also shows that each construct has good internal consistency in terms of ITC.

**Table 3** Reliability and validity of the revised model

| Constructs | Item | Factor loadings (*t*-value) | ITC | CR | AVE | Cronbach's alphas |
|---|---|---|---|---|---|---|
| Requirements uncertainty | RU1 | 0.82 (4.207) | 0.51 | 0.69 | 0.53 | 0.68 |
|  | RU2 | 0.63 (3.955) | 0.51 |  |  |  |
| Top management support | TMS3 | 0.94 (11.569) | 0.74 | 0.81 | 0.60 | 0.80 |
|  | TMS4 | 0.77 (9.377) | 0.67 |  |  |  |
|  | TMS5 | 0.56 (6.604) | 0.52 |  |  |  |
| User participation | UP1 | 0.71 (8.304) | 0.60 | 0.78 | 0.47 | 0.78 |
|  | UP2 | 0.72 (8.368) | 0.60 |  |  |  |
|  | UP3 | 0.65 (7.394) | 0.60 |  |  |  |
|  | UP6 | 0.67 (7.729) | 0.58 |  |  |  |
| Perception gap | PG1 | 0.90 (2.836) | 0.49 | 0.70 | 0.55 | 0.65 |
|  | PG2 | 0.54 (3.040) | 0.49 |  |  |  |

When multiple indicators are used to measure one construct, convergent validity should be assured by examining the reliability of constructs, composite reliability (CR), and the average variance extracted (AVE) [27]. CR is the sum of the loadings squared and then divided by the combination of the sum of the squared loading and the sum of the error terms [4]; it should be greater than 0.6 [44]. Cronbach's alpha is also used to measure construct reliability, and its acceptable threshold is 0.6. Some researchers have indicated that CR is similar to Cronbach's alpha and can be interpreted in the same way [53]. AVE reflects the variance captured by indicators, and should be not less than 0.5 [4]. For each construct, Table 3 shows that the CR (ranging from 0.69 to 0.81) and Cronbach's alpha (ranging from 0.65 to 0.80) both exceed their thresholds. Estimated AVE index values confirm this result for all but one construct. Although the AVE of user participation is lower than 0.5, it is very close to the acceptable value. All in all, the measurement model has good convergent validity.

Finally, we checked discriminant validity, which involves testing whether the measures of constructs are different from each other. It is assessed by testing whether the square root of the AVE is larger than the correlation coefficients [5]. As shown in Table 4, the findings reveal good discriminant validity for all constructs.

**Table 4** Discriminant validity of the measurement model

| Correlations | 1. | 2. | 3. | 4. |
|---|---|---|---|---|
| 1. Requirements uncertainty | 0.73 | | | |
| 2. Top management support | -0.23 | 0.77 | | |
| 3. User participation | -0.15 | 0.18 | 0.69 | |
| 4. Perception gap | 0.27 | -0.24 | -0.13 | 0.74 |

Note: Diagonals represent the square root of the AVE

Notably, the above analysis considers the mean and variance of variables. According to measurement theory, certain statistics, such as mean and variance, are meaningless for ordinal data, and Likert scales produce what is called ordinal data. However, there has been controversy regarding the nature of the data produced by self-reported scales. Although attitudes and feelings cannot be measured with the same precision as pure scientific variables, it is generally accepted in social sciences that self-reported data can be regarded as continuous (interval) and used in parametric statistics for each question item [54]. Additionally, according to the analysis method presented by Byrne [55], if each construct includes several question items, and the Cronbach's alpha score for each construct shows that each construct has good convergent validity, each participant's "average" response for each construct can be calculated. As such, like another work in the literature [4], we used statistics such as standard deviation of the data of Likert scales in this research.

4.2 Structural model

Information about the Squared R ($R^2$) and path coefficients are used to assess the structural model. $R^2$ indicates the amount of variance explained by the independent variables and represents the predictive power of the model [5]. Path coefficients indicate the strength of the relationships between dependent and independent variables. Fig. 1 shows the path analysis results. Hypotheses H1, H2, and H5 were supported at the 0.05 level. However, H3 and H4 were not.

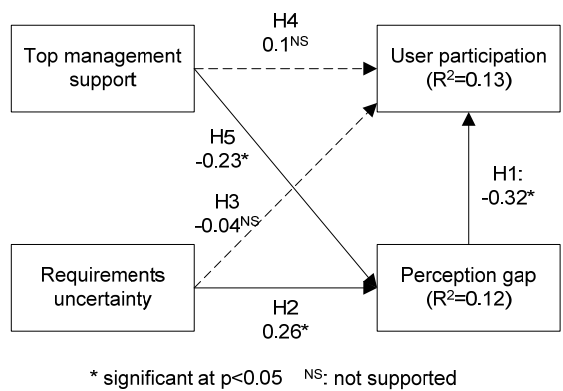

**Fig. 1** The results of path analysis

To test the mediating effects of user-developer perception gap in requirements understanding, supplementary tests were conducted based on the procedures in [56, 57]. Full mediation occurs when the inclusion of the mediating variable negates the effect of the independent variable on the dependent variable [58].

First, the mediating effect of the perception gap between top management support and user participation were examined. The analytical results in Table 5 show that the condition for full mediation was met. Top management support (independent variable) related significantly to perception gap (mediator) with a coefficient value of −0.277, and the perception gap significantly related to user participation (dependent variable) with a coefficient value of −0.359. Without the perception gap, a direct positive relationship between top management support and user participation was established with a coefficient value of 0.198 (p<0.05). When the above three relationships were simultaneously considered in model 4, where perception gap acted as a mediator, however, the relationship between top management support and user participation became non-significant, suggesting that the perception gap fully mediates this relationship.

Next, the mediating effect of the perception gap between requirements uncertainty and user participation was examined. We only examined one model with a direct link between requirements uncertainty and user participation without the mediator of the perception gap and found that the direct link was not confirmed. This condition is contrary to one of the criteria that should be met when a mediator relationship exists [47]. Therefore, perception gap does not have a mediating effect between requirements uncertainty and user participation.

**Table 5** Test results for the mediating effects of perception gap

| | Path | Path coefficient |
|---|---|---|
| Mediating effects between top management support and user participation | | |
| Model 1 | Top management support → user participation | 0.198 * |
| Model 2 | Top management support → perception gap | -0.277 * |
| Model 3 | Perception gap → user participation | -0.359 ** |
| Model 4 | Top management support → user participation | 0.103 ns |
| | Top management support → perception gap | -0.278 * |
| | Perception gap → user participation | -0.33 ** |
| Mediating effects between requirements uncertainty and user participation | | |
| Model 1 | Requirements uncertainty → user participation | -0.156 ns |

* p<0.05; ** p<0.01; ns: not support

**5 Discussions and implications**

During RE, user participation achieves communication between users and developers to facilitate requirements consonance. However, user-developer perception gaps in requirements understanding occur along with user participation. Obviously, perception gaps will cause conflicts and result in negative effect on the projects. Meanwhile, the actual effect of user participation is controversial and not what we had expected [9]. Therefore, this study proposed five hypotheses to examine whether user-developer perception gaps will directly hinder the participation behavior of users. Additionally, the mediating effects of perception gaps were also explored from two aspects: whether it would deteriorate the negative effects of requirements uncertainty on user participation or weaken the positive effects of top management support on user participation.

A survey of 140 subjects indicates that the direct negative influence of perception gaps on user participation does exist. Additionally, consistent with the findings of prior studies, the more requirements uncertainty, the larger the perception gap. The above two relationships show requirements uncertainty indirectly and negatively influence user participation. However, this paper did not support the direct influence of requirements uncertainty on user participation. Therefore, the mediating influence of perception gap between uncertainty and participation did not exist. Surprisingly, and not consistent with the results of prior studies, the fact that top management support has a significant direct effect on user participation was not supported in this paper; this indicates that perception gap fully offsets the positive effect. To be more specific, this paper validated that top management support is helpful for closing the user-developer perception gaps.

5.1 Implications for research

The results of this study have several implications for researchers. This study, to the best of our knowledge, is the first empirical study examining the effects of user-developer perception gaps in requirements understanding on user participation. The study results indicate that perception gap is not only the direct culprit that hinders user participation but also an indirect saboteur that destroys the positive effect of top management support on user participation. On the premise of the fact that user participation promotes project performance, this finding explains the prior conclusion that perception gap has a negative effect on project performance [5]. Therefore, the finding highlights the importance of investigating perception gaps in RE. First, the fact that the positive effect of top management support on user participation is not supported in our paper is not consistent with the prior study of [13] and against common knowledge. The finding is surprising and raises a new research question: Will perception gaps bring other unexpected negative effects that further affect project performance? This is a topic worthy of investigation. Second, we only examined

*Requirements Engineering*, available at: https://link.springer.com/article/10.1007/s00766-017-0266-x, DOI: 10.1007/s0076-017, 2017user-developer perception gaps, but there are actually other gaps between stakeholders. So whether other gaps also bring adverse impacts to a software project needs to be considered. Third, perception gap is essentially a different mind state among stakeholders, and it is related to individual cognition. Thus, applying psychological knowledge into the software engineering domain, especially research about the human aspects, is necessary. Although prior studies have involved psychology [59, 60], the research is still limited. Our research confirmed the necessity of this kind of study. Questions to address in this kind of study include: Do stakeholders with different psychological characteristics, such as personality, exhibit bigger perception gaps than those with the same characteristics? Are there smaller perception gaps between stakeholders with higher cognitive abilities? This research could contribute a new way to close perception gaps. In sum, this study opens new research focuses about perception gaps.

In addition, to facilitate effective user participation, our work presents the need to re-examine the challenges of user participation. First, although our work contributes to finding a new challenge – perception gap – the question is whether there are other challenges that have not been identified. Second, although the challenges in user participation, for example, lack of top management support and involvement conflicts, are listed in [9], the author did not describe which one has the strongest influence. The process of sorting these challenges according to the severity of their influence is worthwhile to explore. Our findings provide some insights on this aspect. From our results, we can infer that the negative impact of perception gaps is greater than that of the lack of top management support, because perception gap fully offsets the positive effect of top management support on user participation. Therefore, we need to explore not only what factors will hinder user participation but also the importance of each factor to fully understand the question of user participation.

5.2 Implications for practice

The results of this paper also have several important implications for software project managers. First, the observed significant negative effect of user-developer perception gaps on user participation implies that project managers should take reasonable actions to close the perception gaps in order to push user participation. It is important to eliminate the direct threat of user participation. Identifying the existence of perception gaps and trying to reduce it should be the prime task of software project managers in RE. Apart from the methods in [4], the support of H5 indicates that top management support is helpful to close the user-developer perception gaps in requirements understanding. In addition, from the viewpoint of practice, many preventive measures are available to assist project managers in reducing the gaps during RE. Training before participation, for example, about software system goal and scope or communication methods, may be useful to guide stakeholders to get consensus on requirements. Additionally, letting users and developers learn each other's characteristics before contact may be useful. Also, from the viewpoint of psychology, people with different personalities may think differently, thus it is easy to see that perception gaps may exist when they exchange ideas. In order to achieve smooth communication with a user and further reduce gaps, the manager could try to use a team member who has a similar personality to that of the user.

Second, requirements uncertainty is always considered the number one enemy by software project managers. Bano and Zowghi [9] argued that uncertainty was one of the challenges of user participation. But the finding that H3 was not supported indicates that users would not reject participation due only to requirements uncertainty. The finding is actually in line with the reality, because some requirements uncertainty is caused by users. However, the fact that requirements uncertainty exacerbates the perception gap is validated in this paper. Therefore, software managers should pay attention to the direct negative influence of requirements uncertainty instead of worrying about requirements uncertainty. Third, to achieve the effect of user participation, the manager should give a different priority to each challenge of user participation when managing the challenges.

**6 Conclusions**

This study has attempted to examine not only the direct influence of user-developer perception gaps in requirements understanding on user participation, but also its mediating roles by combining top management support and requirements uncertainty because they represent two different aspects (human behavior and non-human matter) of the external environment. The five relationships between the four variables were hypothesized and tested. Three hypotheses were empirically supported and two were not. The research results show that user-developer perception gaps in requirements understanding not only have a direct negative effect, but also totally offset the positive influence of top management support on user participation, although perception gaps do not have a mediating role between requirements uncertainty and user participation.

This paper contributes to several conclusions that have not been studied before. First, this paper empirically validated that perception gaps in requirements understanding have a direct negative effect on user participation. Also notable was the finding that perception gaps in requirements understanding totally offset the positive influence of top management support on user participation, which confirms the mediating effect of perception gaps on user participation. Combining with prior knowledge from the literature, this help us to understand the intertwined relationships between the attributes. For one thing, the findings serve to reveal the danger of perception gaps and the paper contributes a new way – top management support – to close the gap. In addition, the findings contribute to helping us understand the problems that user participation faces and improve it in practice by reducing the user-developer perception gap.

However, limitations that pertain to our study need to be acknowledged as our results are bounded by threats to validity. Our acknowledgement of these limitations also suggests new directions for future studies. The first is related to the sample frame, which is based on part-time students. Because there is no evidence that the sample frame is a typical representation, the sample data may affect the validity of the results. The number of top managers is relatively low, which poses a threat to the validity of results related to top management. In addition, we focused on examining perception gaps in requirements understanding. Although we know that most subjects whose roles are tester, developer, or others in the survey, they have done or are doing activities related to RE, we must acknowledge that there may be some subjects who know little about RE. Thus, the results limitations should be taken with this consideration in mind. As such, investigating and comparing different samples to ensure the validity should be done in the future. Second, the responses to questions in the survey are all subjective, which leads to the subjectivity of the results. Thus, we should try to measure some variables objectively to reduce the subjectivity threat in the future. The third limitation is related to the research focus: only the user-developer perception gap is considered in this paper. Future studies should take other stakeholder perception gaps into account, because they also exist in a software project. Fourth, the final goal of a project is its success. Since user participation is also expected to increase project success, project performance could be related to a perception gap in the future. Finally, the insignificant direct effect of top management support on user participation found in this work is rather surprising. Due to the mediating role of the perception gap, the finding is different from previous studies. We, thus, encourage future research that aims to investigate the effects of top management support on user participation in a variety of contexts by taking into consideration other potential mediating factors.

**Acknowledgements**

The authors thank the survey respondents for providing valuable data. This research was supported by the China Scholarship Council (CSC).

**Appendix A. Measurements of constructs**

| Item | Questions | Reference |
|---|---|---|
| Requirements uncertainty (RU) | | |
| RU1 | There are lots of new requirements. | [15, 40] |
| RU2 | Many requirements identified at the beginning need to be modified after the stage of requirements elicitation. | |
| *RU3* | *Most of the requirements identified at the beginning were deleted in later phases. (discarded)* | |
| *RU4* | *Requirements specification was approved only through more than one inspection. (discarded)* | |
| *RU5* | *There are untestable requirements in the requirements specification. (discarded)* | |
| Top management support (TMS) | | |
| *TMS1* | *Top management is interested in the requirements elicitation. (discarded)* | [38, 44-46] |
| *TMS2* | *Top management believes the project is important and beneficial to the organization. (discarded)* | |
| TMS3 | Top management often attends the project meetings. | |
| TMS4 | Top management often hosts the project meetings. | |

| TMS5 | Top management solves project problems in time. | |
|---|---|---|
| *TMS6* | *Top management is eager to monitor and learn project progress. (discarded)* | |
| User participation (UP) | | |
| UP1 | The actual time of user participation during project start-up phase. | [41-43] |
| UP2 | The actual time of user participation during defining project goals. | |
| UP3 | The actual time of user participation during defining user requirements. | |
| *UP4* | *The actual time of user participation during describing business information process. (discarded)* | |
| *UP5* | *The actual time of user participation during defining the inputs and outputs of software system. (discarded)* | |
| UP6 | Sum up, the actual time of user participation during project requirements phase. | |
| Perception gap (PG) | | |
| PG1 | There was a clear perception gap between users and developers understanding of the requirements specification. | [4] |
| PG2 | There was a clear perception gap between users and developers understanding of the project scope/objectives. | |
| *PG3* | *There was a clear perception gap between users and developers understanding of the project success criteria. (discarded)* | |
| *PG4* | *There was a clear perception gap between users and developers understanding of the system capabilities and limitations. (discarded)* | |
| *PG5* | *There was a clear perception gap between users and developers understanding of the inputs and outputs of the system. (discarded)* | |

**Appendix B. Characteristics of the organizations and projects of respondents**

| Variables | Categories | Number | Percent |
|---|---|---|---|
| Average team size | <=7 people | 62 | 44.3 |
| | 8-15 | 43 | 31 |
| | 16-25 | 17 | 12 |
| | >=26 | 17 | 12 |
| | Missing | 1 | 0.7 |
| Average project duration | <= 1years | 69 | 49.3 |
| | (1, 2] years | 53 | 37.9 |
| | (2, 3] years | 10 | 7.1 |
| | (3, 5]years | 3 | 2.1 |
| | > 5 years | 4 | 2.9 |
| | Missing | 1 | 0.7 |
| Average software provider company size | <= 10 people | 14 | 10 |
| | 11-50 people | 30 | 21.4 |
| | 51-100 people | 32 | 22.9 |
| | 101-200 people | 15 | 10.7 |
| | 201-500 people | 15 | 10.7 |
| | > 500 people | 30 | 21.4 |
| Length of time the software provider company has been established | Missing | 4 | 2.9 |
| | <= 1years | 9 | 6.4 |
| | (1, 3] years | 29 | 21 |
| | (3, 5]years | 24 | 17 |
| | (5, 10]years | 44 | 31.4 |
| | (10, 20] years | 23 | 16.4 |
| | >20 years | 8 | 5.7 |
| | Missing | 3 | 2.1 |
| Industry type of software service | Manufacturing | 10 | 7.1 |
| | Transportation | 7 | 5 |
| | Petroleum & chemical | 9 | 6.4 |
| | Information technology (IT) | 15 | 10.7 |
| | Food & clothing | 3 | 2.1 |
| | Cultural and sports industry | 3 | 2.1 |
| | Construction | 7 | 5 |
| | Biological/medicine | 9 | 6.4 |
| | Retail & wholesale | 6 | 4.3 |
| | Financial | 15 | 10.7 |
| | Government | 25 | 18 |
| | Education | 19 | 13.6 |
| | Others | 12 | 8.6 |